\documentclass[]{pasj02}
\usepackage[switch,mathlines]{lineno} % add line number to manuscript
\usepackage{multirow}
\usepackage{threeparttable}

\jyear{2025}
\Received{}
\Accepted{}%{yyyy/mm/dd}
\begin{document}

\title{Benchmark of the Fe \emissiontype{XXV} $\mathcal{R}$ ratio in photoionized plasma during
eclipse of Centaurus X-3 with XRISM/\textit{Resolve}}

\author{
 Yuto \textsc{Mochizuki},\altaffilmark{1,2}\altemailmark\orcid{0000-0003-3224-1821}
 \email{mochizuki-yuto652@g.ecc.u-tokyo.ac.jp, mochizuki@ac.jaxa.jp}
 Masahiro \textsc{Tsujimoto},\altaffilmark{1,2}\orcid{0000-0002-9184-5556}
 Maurice A. \textsc{Leutenegger},\altaffilmark{3}\orcid{0000-0002-3331-7595}
 Liyi~\textsc{Gu},\altaffilmark{4}\orcid{0000-0001-9911-7038}
 Ralf~\textsc{Ballhausen},\altaffilmark{3,5}\orcid{0000-0002-1118-8470}
 Ehud~\textsc{Behar},\altaffilmark{6}\orcid{0000-0001-9735-4873}
 Paul A. \textsc{Draghis},\altaffilmark{7}\orcid{0000-0002-2218-2306}
 Natalie~\textsc{Hell},\altaffilmark{8}\orcid{0000-0003-3057-1536}
 \and
 Pragati~\textsc{Pradhan}\altaffilmark{9}\orcid{0000-0002-1131-3059}
}
 \altaffiltext{1}{Department of Astronomy, Graduate School of Science, The University of
 Tokyo, Bunkyo-ku, Tokyo 113-0033, Japan}
 \altaffiltext{2}{Japan Aerospace Exploration Agency, Institute of Space and
 Astronautical Science, Sagamihara, Kanagawa 252-5210,
 Japan}
 \altaffiltext{3}{NASA/Goddard Space Flight Center, Greenbelt, MD 20771, USA}
 \altaffiltext{4}{SRON Netherlands Institute for Space Research, Niels Bohrweg 4, 2333
 CA Leiden, The Netherlands}
 \altaffiltext{5}{Department of Astronomy, University of Maryland, College Park, MD 20742, USA}
 \altaffiltext{6}{Physics Department, Technion, Haifa 32000, Israel}
 \altaffiltext{7}{MIT Kavli Institute for Astrophysics and Space Research, Massachusetts Institute of Technology, 70 Vassar St,
Cambridge, MA 02139, USA}
 \altaffiltext{8}{Lawrence Livermore National Laboratory, Livermore, CA 94550, USA}
 \altaffiltext{9}{Department of Physics, Embry-Riddle Aeronautical University: Prescott, AZ 86301, USA}
\maketitle

\begin{abstract}
 The $\mathcal{R}$ ratio is a useful diagnostic of the X-ray emitting astrophysical
 plasmas defined as the intensity ratio of the forbidden over the inter-combination
 lines in the K$\alpha$ line complex of He-like ions. The value is altered by excitation
 processes (electron impact or UV photoexcitation) from the metastable upper level of
 the forbidden line, thereby constraining the electron density or UV field
 intensity. The diagnostic has been applied mostly in electron density constraints in
 collisionally ionized plasmas using low-$Z$ elements as was originally proposed for the
 Sun (Gabriel \& Jordan (\yearcite{gabriel1969}, MNRAS, 145, 241)), but it can also be used in photoionized plasmas.  To make use
 of this diagnostic, we need to know its value in the limit of no excitation of
 metastables ($\mathcal{R}_{0}$), which depends on the element, how the plasmas are
 formed, how the lines are propagated, and the spectral resolution affecting line
 blending principally with satellite lines from Li-like ions. We benchmark
 $\mathcal{R}_0$ for photoionized plasmas by comparing calculations using radiative
 transfer codes and observation data taken with the \textit{Resolve} X-ray
 microcalorimter onboard XRISM. We use the Fe\emissiontype{XXV} He$\alpha$ line complex
 of the photo-ionized plasma in Centaurus X-3 observed during eclipse, in which the
 plasma is expected to be in the limit of no metastable excitation. The measured
 $\mathcal{R} = 0.65 \pm 0.08$ is consistent with the value calculated using
 \texttt{xstar} for the plasma parameters derived from other line ratios of the
 spectrum. We conclude that the $\mathcal{R}$ ratio diagnostic can be used for high-$Z$
 elements such as Fe in photoionized plasmas, which has wide applications in plasmas
 around compact objects at various scales.
\end{abstract}

% \pagewiselinenumbers

\section{Introduction}\label{s1}

Line-driven winds are observed in various astrophysical phenomena such as the mass
loss in early-type stars \citep{morton1967,hutchings1976,lamers1976,abbott1978}, disk
winds in cataclysmic variables \citep{heap1978,cordova1982}, and possibly ultra-fast
($\sim 0.1c$) outflows in narrow-line Seyfert 1 galaxies
\citep{tombesi2010,tombesi2013}.  These winds are primarily accelerated by transferring
momentum from UV photons to matter via photoexcitation of mildly ionized (Li- to
Ne-like) ions \citep{castor1975,proga2000,nomura2016}.  The representative agents
\citep{lucy1970} include 2p\,$^{2}P_{1/2,3/2} \rightarrow$2s\,$^{2}S_{1/2}$ transition
of C\emissiontype{IV} at 1550 and 1548~\AA{} (8.00 and 8.01~eV), N\emissiontype{V} at
1243 and 1239~\AA{} (9.97 and 10.0~eV), and O\emissiontype{VI} at 1038 and 1032~\AA{}
(11.9 and 12.0~eV).  The spectral energy distribution (SED) of the incident source in
the UV range is thus important to quantify the dynamics of this ubiquitous mechanism,
yet its observational access is often hampered by the extinction in the line of sight
\citep{kalberla2005}, in particular, for active galactic nuclei (AGNs).

An interesting idea was proposed by \citet{porter2007b}, in which X-ray spectra can be
used to constrain the local UV flux of AGNs. Conspicuous X-ray lines are due to $n=2
\rightarrow 1$ transitions of highly-ionized (H- and He-like) ions.  Their energies are
in the X-ray range (0.1--10~keV), while the energy differences between different
sublevels within $n=2$ are in the UV range (200--2000~\AA;
\cite{morton1991,artemyev2005,yerokhin2019}). The intensity ratios among the $n=2
\rightarrow 1$ X-ray lines of He-like ions are affected by photoexcitation by local UV
flux, among other influences. We can thus constrain the local UV flux through X-ray
spectra that are less attenuated by the extinction in the line of sight.

The principal UV flux diagnostic is the $\mathcal{R}$ ratio in the He-like $n=2
\rightarrow 1$ (hereafter, He$\alpha$) line complex. It is defined as the ratio between
the forbidden ($f$) and inter-combination ($i$) lines. The diagnostic was originally
proposed primarily as a probe of the electron density in collisionally-ionized plasmas
\citep{gabriel1969}, as electrons are another agent to excite between the $n=2$
sublevels. There are many applications of this electron density diagnostic using low-$Z$
elements ($Z \lesssim 18$, where $Z$ is the atomic number) in collisionally-ionized
plasmas in the Sun
\citep{doschek1980,dubau1981,feldman1980,parmar1981,tanaka1982,tanaka1986,watanabe2024},
extra-solar stars \citep{canizares2000,ness2001,ness2002,audard2003,nordon2007}, and
cataclysmic variable
\citep{mukai2001,mauche2002,mauche2003,vrielmann2005,itoh2006,schlegel2014}.

The application of the $\mathcal{R}$ ratio diagnostic can be extended (i) for the local
UV flux constraints as opposed to the electron density constraints or a combination of
them, (ii) for plasmas at the photo-ionization equilibrium (PIE) as opposed to those at
the collisional-ionization equilibrium (CIE), and (iii) for the use of high-$Z$ elements as
opposed to low-$Z$ elements. Many efforts have been made to refine the theory
\citep{blumenthal1972,mewe1978,pradhan1981,porquet2000a,bautista2000,coupe2004} and the
observational demonstration in actual use cases. For (i), the most successful case is in
early-type stars, in which the ratio is altered by the local UV flux from the OB stars
in the CIE plasmas
\citep{kahn2001,cassinelli2001,waldron2001,wojdowski2001,leutenegger2006,cohen2022}.
For (ii), warm absorbers in AGNs and stellar winds in X-ray binaries are among the
representative use cases \citep{sako2000,kinkhabwala2002,lomaeva2020}.  For (iii), the
actual case is limited, as the existing dispersive X-ray spectrometers have insufficient
spectral resolution for high-$Z$ elements.

In all these applications, we need to have tools to calculate the $\mathcal{R}$ value at
the limit of no radiative and collisional excitation ($\mathcal{R}_0$) in various
circumstances. In this paper, we aim to benchmark this for a combination that has not been
explored enough before ---local UV flux constraint in a PIE plasma using a high-$Z$
element. We use the X-ray spectra of an eclipsing high-mass X-ray binary (HMXB)
---Centaurus X-3 (Cen X-3)--- taken with the X-ray microcalorimeter spectrometer
\textit{Resolve} \citep{ishisaki2022} onboard the XRISM (X-Ray Imaging and Spectroscopy
Mission; \cite{tashiro2025}) satellite.

HMXBs are an interesting laboratory for the verification of the diagnostic. They have
the local X-ray flux from the compact object to generate PIE plasmas for the He$\alpha$
emission of high-$Z$ elements. They also have local UV flux from the O star, the stellar
surface of the compact object if it is a neutron star (NS), and the accretion disk
around the compact object. Some of them, including Cen X-3, exhibit total
eclipses. During eclipses, we can extract the He$\alpha$ line emission not contaminated
by the incident continuum emission with absorption lines \citep{pradhan2024a}, which would otherwise dominate
the observed spectra and offset the emission line intensities \citep{mehdipour2015}.
X-ray microcalorimeter spectrometers allow the resolution of the He$\alpha$ complex of
high-$Z$ elements such as Fe\emissiontype{XXV} into fine-structure and satellite
lines. A verification in such a system will pave the way for the application of a wide
range of systems including AGNs and low-mass X-ray binaries.

We start with a brief overview of the $\mathcal{R}$ ratio diagnostics (section~\ref{s2}) with
particular emphasis on the important differences in physics relevant in the application
of this study as opposed to the others. We then present the data to apply the diagnostic
(section~\ref{s3}) and radiative transfer calculation (section~\ref{s4}) to interpret the
ratio. Finally, we combine these two to give constraints to the local UV flux, thus the
location of the X-ray emitting plasmas in the system (section~\ref{s5}) before concluding in
section~\ref{s6}.

\section{Overview of the $\mathcal{R}$ ratio}\label{s2}

\subsection{Notations}\label{s2-1}

A comprehensive review of the $\mathcal{R}$ ratio diagnostic is given in
\citet{porquet2010}. Here, we present only the notation that will be used in this
paper. First, the ratio is given by
\begin{equation}
 \label{e01}
 \mathcal{R} \equiv \frac{I_{\mathrm{z}}}{I_{\mathrm{x}}+I_{\mathrm{y}}},
\end{equation}
where $I_{\mathrm{x}}$, $I_{\mathrm{y}}$, and $I_{\mathrm{z}}$ are the line flux
(erg~s$^{-1}$) of the $x$, $y$, and $z$ lines in the He$\alpha$ complex
\citep{gabriel1969}. The $x$ and $y$ lines are the inter-combination lines respectively
from the 1s2p\,$^{3}P_{2}$ and 1s2p\,$^{3}P_{1}$ level to the ground
1s$^{2}$\,$^{1}S_{0}$ level of the He-like ion, while $z$ is the forbidden line from the
metastable level 1s2s\,$^{3}S_{1}$ to ground \citep{gabriel1972}.

The formation processes of these lines can be better understood by comparing them with
those of other lines. The $w$ line is the resonance line from the 1s2p\,$^{1}P_{1}$
level to the ground. These four main lines of He-like ions ($w$, $x$, $y$, and $z$) are
accompanied by dozens of satellite lines of Li-like ions ($a$--$v$), which are caused by
the $n=2 \rightarrow 1$ transitions with a slightly different energy due to the presence
of a spectator electron in the $n=2$ shell \citep{gabriel1969b}.  We used two
representative lines that are predominantly formed by different processes. The $q$ line
is the transition from 1s2s($^{3}S$)2p\,$^{2}P_{3/2}$ to the ground level
2s\,$^{2}S_{1/2}$ of the Li-like ions. The $j$ line is the transition from
1s2p$^{2}$\,$^{2}D_{5/2}$ to 2p\,$^{2}P_{3/2}$. For convenience, we use, for example,
$\overline{z}$ and $\underline{z}$ to indicate the upper and lower levels of the $z$
line, respectively.

The $\mathcal{R}$ ratio changes as a function of the plasma density or the local UV
photon flux. The fact that the $\overline{z}$ level is a metastable state is
utilized. Before decaying radiatively to $\underline{z}$, $\overline{z}$ is excited to
$\overline{x}$ or $\overline{y}$ (or to the 1s2p\,$^{3}P_{0}$ level, which however does not
radiatively decay to the ground level, hence is not included in the $\mathcal{R}$ ratio) either
collisionally or radiatively. In case of collision, the collision partner is mostly
electrons, but protons and alpha particles contribute to some extent
\citep{pradhan1981}. They can be important even in PIE plasmas, as the energy required
for excitation is small (5--50~eV). In case of radiation, photons that matter are in the
UV range (200--2000~\AA). Both influences tend to come together in actual cases
\citep{peretz2019,saathoff2024}. The $\mathcal{R}$ ratio is altered as
\begin{equation}
 \label{e02}
 \mathcal{R} = \mathcal{R}_{0} \frac{1}{1+\left(\frac{n_{\mathrm{e}}}{n_{\mathrm{crit}}}\right)+\left(\frac{\phi_{\mathrm{UV}}}{\phi_{\mathrm{crit}}}\right)},
\end{equation}
where $n_{\mathrm{crit}}$ is the critical density at which the collisional excitation from
$\overline{z}$ to $\overline{x}$ or $\overline{y}$ equals to the radiative de-excitation
from $\overline{z}$ to $\underline{z}$. Likewise, $\phi_{\mathrm{crit}}$ is the critical
photon flux at which the photoexcitation from $\overline{z}$ to $\overline{x}$ or
$\overline{y}$ equals to the radiative deexcitation from $\overline{z}$ to
$\underline{z}$.  The measured $\mathcal{R}$ translates into the electron density
$n_{\mathrm{e}}$ or the UV photon flux $\phi_{\mathrm{UV}}$ if $\mathcal{R}_{0}$ is
known and the $n_{\mathrm{e}}$ or $\phi_{\mathrm{UV}}$ are sufficiently below their
critical values, respectively.

\subsection{Important differences}\label{s2-2}
Before applying the $\mathcal{R}$ ratio diagnostic to local UV flux constraint in a PIE
plasma using high-$Z$ element with an X-ray microcalorimeter specrometer, we need to be
aware of several important differences from its original use for the electron density
constraint in CIE plasmas using low-$Z$ elements with the X-ray grating
spectrometers. They are discussed in \citet{porquet2010}. We list three of them relevant
for this work in detail below (sub-subsection~\ref{s2-2-1}--\ref{s2-2-3}).

\begin{figure*}[htpt!]
 \begin{center}
  \includegraphics[width=1.0\columnwidth]{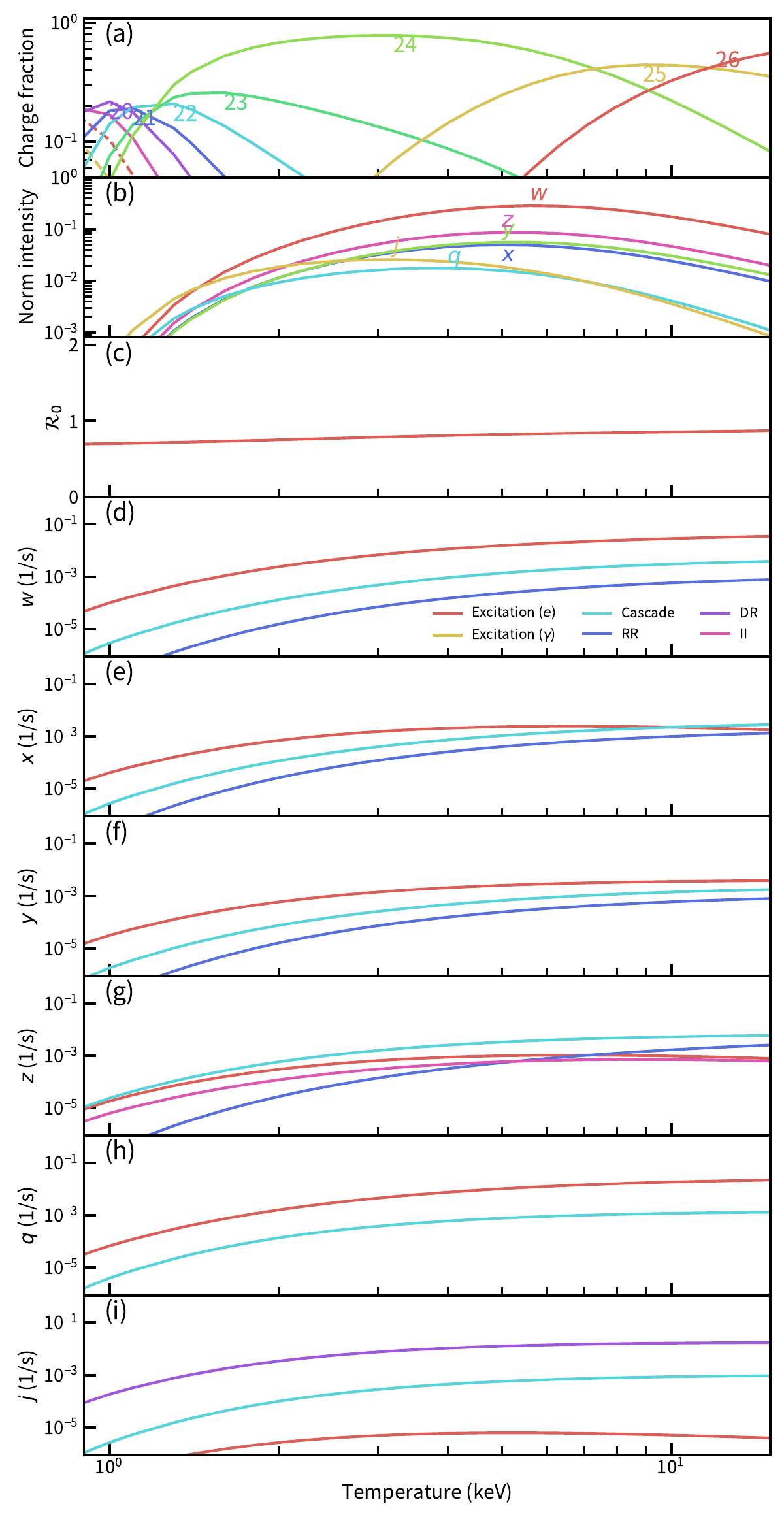}
  \includegraphics[width=1.0\columnwidth]{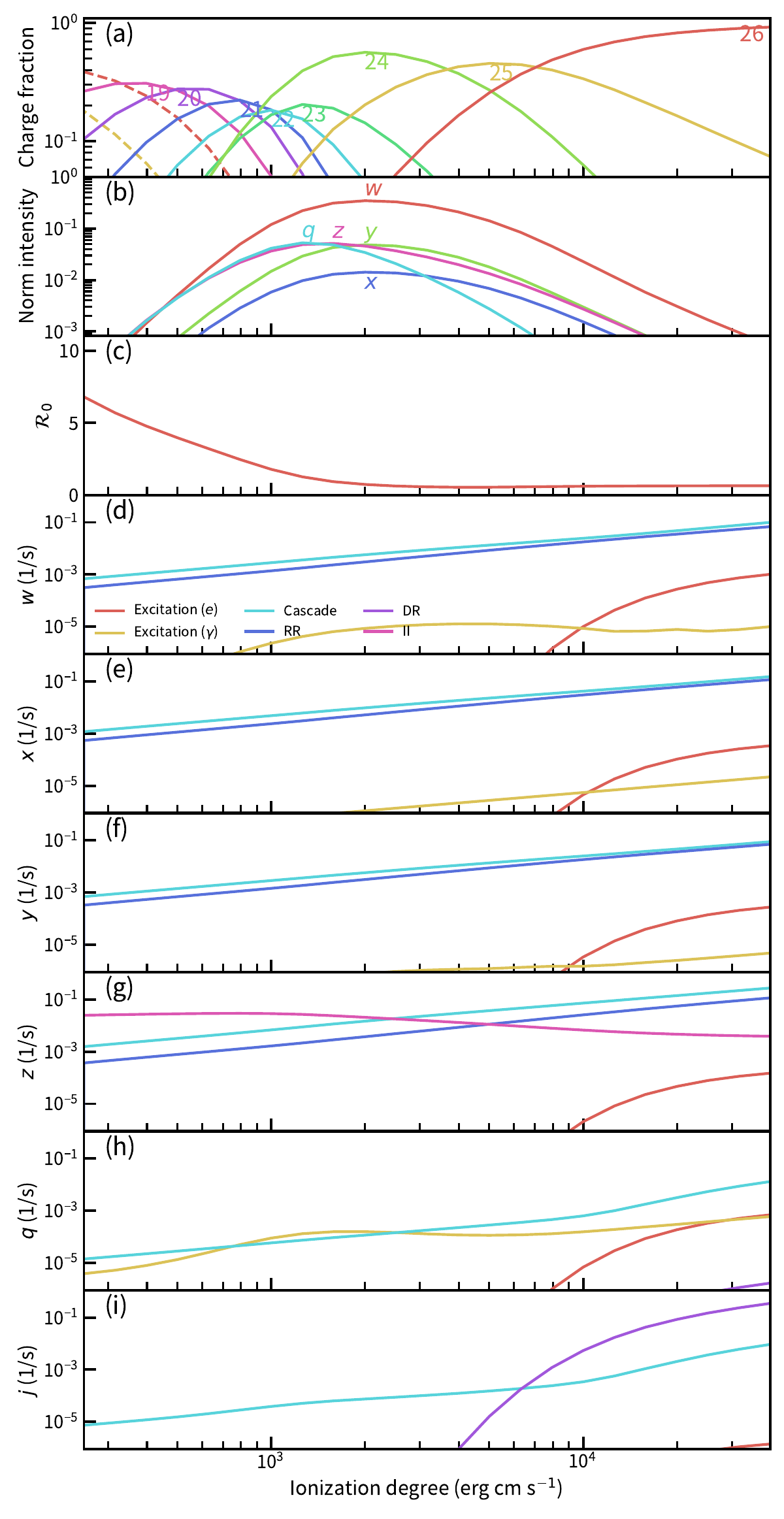}
 \end{center}
 \caption{Fe lines of the CIE (left) and PIE (right) plasmas at the low-density
 (10$^{10}$~cm$^{-3}$) and optically-thin limit as a function of the plasma temperature
 (CIE) or the ionization degree (PIE). At these limits, $\mathcal{R}$ is independent of
 the density and RT effects are negligible for the lines. For PIE, the power-law
 incident emission was assumed with a photon index of 2 and the electron temperature was
 determined by calculating the thermal balance.
 (a) Charge state distribution of Fe. The labels indicate the number of stripped
 electrons in the ions. (b) Normalized intensity of the $w$, $x$, $y$, $z$, $j$, and $q$
 lines, (c) $\mathcal{R}_0$, (d)---(i) Rate of major processes to populate the upper
 levels of $w$, $x$, $y$, $z$, $q$, and $j$. The processes are the excitation from the
 ground state collisionally by the electrons ($e$) or radiatively by photos ($\gamma$),
 cascade from higher levels of He-like ions, direct radiative or dielectronic
 recombination (RR or DR) of the H-like ions, and inner-shell ionization (II) of the
 Li-like ions.
 {Alt text: charge state distribution, line intensity, and atomic processes to populate
 their upper levels calculated using spex.}
}
 \label{f01}
\end{figure*}

\subsubsection{$\mathcal{R}_0$ value}\label{s2-2-1}
The value of $\mathcal{R}_0$, which is a pre-requisite for the $\mathcal{R}$ ratio
diagnostic, depends on $Z$ and whether the plasma is CIE or PIE. This is because the
relative importance of the atomic processes to populate $\overline{x}$, $\overline{y}$,
and $\overline{z}$ is different. Figure~\ref{f01} shows (a) the Fe charge distribution,
(b) the intensity of the $w$, $x$, $y$, $z$, $j$, and $q$ lines, (c) $\mathcal{R}_0$,
and (d--i) the atomic processes to populate the upper levels of these lines as a
function of the plasma temperature for the CIE plasma (left) or as a function of the
ionization degree ($\xi_{\mathrm{X}}$ erg~cm~s$^{-1}$) for the PIE plasma (right). Here,
$\xi_{X} \equiv \frac{L_{\mathrm{X}}}{n_{\mathrm{e}}r^{2}}$, in which $L_{\mathrm{X}}$
is the luminosity integrated over 1--1000~Ryd, $n_{\mathrm{e}}$ is the electron density,
and $r$ is the distance from the X-ray source to the illuminated surface of the plasma
\citep{tarter1969}.  In both cases, the calculation was made using the \texttt{spex}
code version 3.08.01 \citep{kaastra1996} at the low-density and the optically-thin
limit, utilizing its feature to dump the rates of individual physical processes of line
formation.

In CIE plasmas (figure~\ref{f01} left), the dominant process to populate $\overline{w}$,
$\overline{x}$, and $\overline{y}$ is the electron collision excitation from the ground
state of the He-like ions, while that for $\overline{z}$ is the cascade for all
temperature ranges shown in the figure. The $\mathcal{R}_{0}$ value is almost constant
against the varying temperature, which makes the $\mathcal{R}$ ratio useful as
deviations from $\mathcal{R}_0$. The $\overline{q}$ level is also populated
predominantly by electron collision excitation of Li-like ions, whereas the
$\overline{j}$ level is populated by dielectronic recombination (DR) of He-like ions.

In PIE plasmas (figure~\ref{f01} right), collisional processes can still contribute to
the level populations, especially for the low-energy, intra-shell transitions, i.e.,
$n=2 \rightarrow 2$, relevant for the $\mathcal{R}$ ratio diagnostic discussed here.
However, in the low-density regime, especially at densities below the critical density
$n_{\mathrm{crit}}$, collisions are less important in determining the upper level
population of the He$\alpha$ lines.  The critical density ranges from $\sim 10^7$
cm$^{-3}$ for $Z=6$ (C) \citep{porquet2000a} to $10^{16}$ cm$^{-3}$ for $Z=26$ (Fe)
\citep{blumenthal1972}. Instead, the cascade of the He-like ion becomes more important
to populate $\overline{w}$, $\overline{x}$, and $\overline{y}$. For the $\overline{z}$
level, the cascade dominates at $\log{\xi_{\mathrm{X}}} \gtrsim 3.5$, while the
inner-shell ionization of the Li-like ions dominates at $\log{\xi_{\mathrm{X}}} \lesssim
3.5$.
Because of this, the
Fe\emissiontype{XXV} $\mathcal{R}$ ratio diagnostic is only useful in
$\log{\xi_{\mathrm{X}}} \gtrsim 3.5$, where $\mathcal{R}_0$ is nearly constant against
$\xi_{\mathrm{X}}$.
For the $\overline{q}$ level, the recombination cascade dominates at
$\log{\xi_{\mathrm{X}}} \gtrsim 3$, while the inner-shell photo-excitation of the
Li-like ion dominates at $\log{\xi_{\mathrm{X}}} \lesssim 3$.

\subsubsection{Radiative transfer effects}\label{s2-2-2}

In the PIE application of the $\mathcal{R}$ ratio diagnostic, it is important to take
the radiative transfer (RT) effects into account
\citep{bautista2000,coupe2004,porquet2000a}. In practical use cases of AGNs and X-ray
binaries, emission lines of a large oscillator strength ($f_{\mathrm{osc}}$) can be
optically thick, such as the Fe\emissiontype{XXV} $w$ line
($f_{\mathrm{osc}}=0.72$). Photons at the transition energy may undergo multiple
scattering before escaping from the system, through which the photon propagation
direction changes. The line intensity therefore has a strong geometrical dependence and
is stronger when the continuum source with the corresponding absorption line is not
observed in the beam. In fact, this was observed in the X-ray grating spectra of Cen X-3
\citep{wojdowski2003}.

Fortunately, the $\mathcal{R}$ ratio is composed of the lines with a small
$f_{\mathrm{osc}}$ value of $1.7\times 10^{-5}$ ($x$), $5.8 \times 10^{-2}$ ($y$), and
$3.0\times 10^{-7}$ ($z$) for Fe\emissiontype{XXV}. The RT effects are less pronounced
compared to other diagnostics using the $w$ line \citep{coupe2004}. Still, the line
intensities depend on the charge and level populations, which deviate from the local
thermodynamic equilibrium condition because of the strong radiation field from the
central compact source. RT calculations are mandatory, and we will present some in
section~\ref{s4}.

\subsubsection{Spectral resolution of spectrometers}\label{s2-2-3}

The application of the $\mathcal{R}$ ratio diagnostic has been limited to low-$Z$
elements such as N\emissiontype{VI}, O\emissiontype{VII}, Ne\emissiontype{IX},
Mg\emissiontype{XI}, Si\emissiontype{XIII}, and S\emissiontype{XV}. They are in the soft
energy band ($\lesssim$4~keV) accessible with dispersive X-ray spectrometers such as
the low and high energy transmission grating spectrometer (LETGS; \cite{brinkman2000a}
and HETGS; \cite{canizares2005}) onboard the Chandra X-ray observatory and the
reflection grating spectrometer (RGS; \cite{denherder2001}) onboard the XMM-Newton
observatory. These spectrometers cannot resolve fine-structure levels that differ
only in the total angular momentum $J$ such as the $x$ and $y$ lines or the satellite
lines. Therefore, contamination by blending must be considered for proper
use of the diagnostic \citep{porquet2010}.

With the X-ray microcalorimeter spectrometer \textit{Resolve} onboard XRISM, we can
access high-$Z$ elements such as Ar\emissiontype{XVII}, Ca\emissiontype{XIX},
Fe\emissiontype{XXV}, and Ni\emissiontype{XXVII} in the hard energy band
($\gtrsim$4~keV).  Its spectral resolution allows us to resolve the fine-structure
levels of $x$ and $y$ for $Z>20$ (Ca). The relative contribution of the satellite lines
increases as $Z$ increases, but they can be resolved to some extent from the main lines
with \textit{Resolve} \citep{kurihara2025,tsujimoto2025a}.

\section{Data}\label{s3}

\subsection{Observation \& Data Reduction}\label{s3-1}

Cen X-3 is an eclipsing HMXB consisting of an O6--8III star \citep{krzeminski1974} and a
NS \citep{schreier1972} with an orbital period of 2.08 days
\citep{hutchings1979,falanga2015}.
In this system, the semimajor axis is $a=42.1$ light-seconds \citep{bildsten1997} and the eccentricity is $e=0.00$ \citep{kelley1983}.
The NS is totally eclipsed by the O star during
$\sim$20\% of the orbit. It is one of the brightest sources of this class, with X-ray
flux predominantly from the NS of $\approx 10^{-9}$ erg~s$^{-1}$~cm$^{-2}$ out of the
eclipse, which decreases to $\sim 10^{-10}$ erg~s$^{-1}$~cm$^{-2}$ during the eclipse
\citep{torregrosa2022}. The Fe \emissiontype{XXV} He$\alpha$ line complex is observed
both in and out of the eclipse \citep{wojdowski2001,wojdowski2003}.

We use the \textit{Resolve} instrument onboard XRISM. \textit{Resolve} hosts an array of
6$\times$6 X-ray microcalorimeter pixels placed at the focus of an X-ray telescope,
which serves as a non-dispersive, high throughput, and high-resolution X-ray
spectrograph with an energy resolution of $E/\Delta E \sim 1400$ (FWHM at 5.9~keV;
\cite{porter2024}), and an effective area of$\sim 174$ cm$^2$ at 6~keV with the gate valve
closed \citep{hayashi2024}. Details of the instrument can be found in \citet{sato2023}
and the references therein.

The observation was made on 2024 February 12–15 (sequence number 300003010) as a part of
the performance verification program. A total of 196~ks exposure covered 1.2 orbits with
a nearly continuous coverage. Details of the data set are given in
\citet{mochizuki2024b}. The beginning and end of the observation were taken during two
subsequent eclipses. We combined the two parts with a total of 41.4~ks exposure. The
average count rate during eclipse was 5~s$^{-1}$, and most of the events are high primary (Hp) events
suitable for high-resolution spectroscopy.

We retrieved pipeline products with processing version 03.00.011.008. We applied
standard screening, removed events other than the Hp grade, and applied additional
screening \citep{mochizuki2025}. We used
\texttt{HEASoft} version 6.34 for data reduction and \texttt{Xspec} version 12.14.1
\citep{xspec} for spectral fitting. The errors quoted hereafter indicate a 1$\sigma$
statistical uncertainty.

\subsection{Data Analysis}\label{s3-2}

\begin{figure}[htpt!]
 \begin{center}
  \includegraphics[width=1.0\columnwidth]{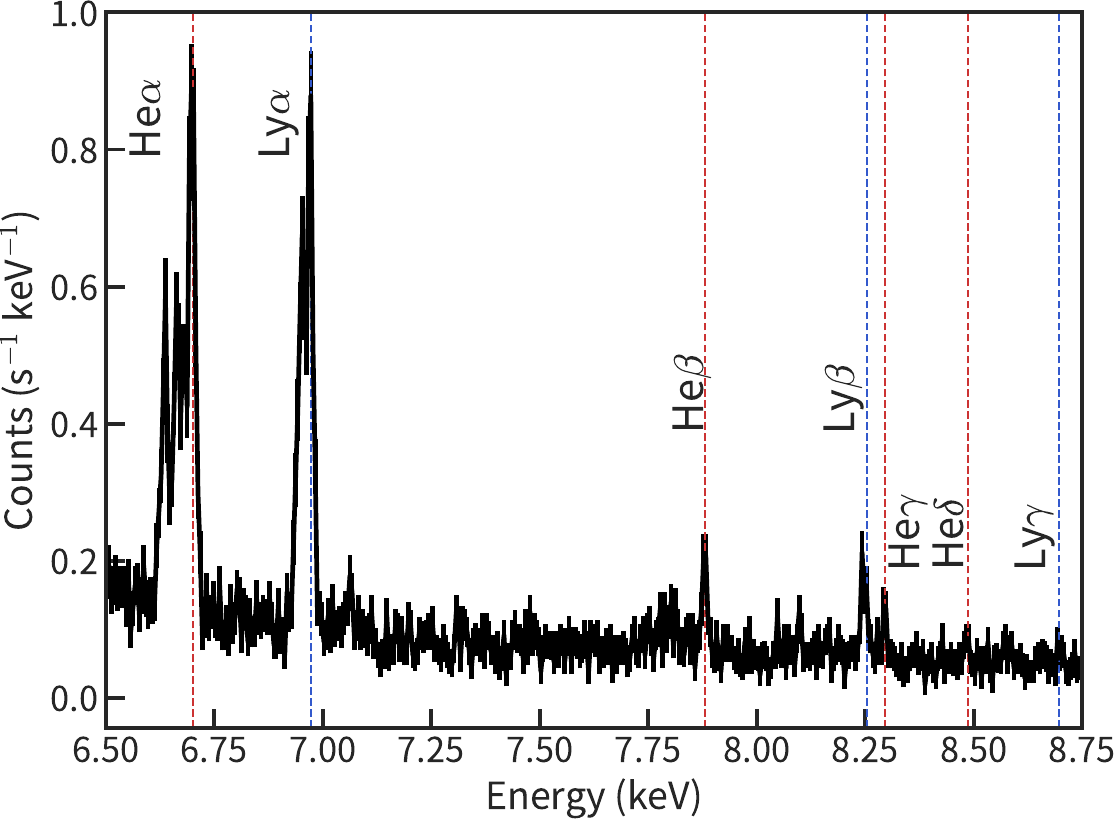}
 \end{center}
 \caption{Fe K-band spectrum during the eclipses taken with \textit{Resolve}.  Detected
 emission lines of Fe \emissiontype{XXV} He series and Fe \emissiontype{XXVI} Lyman
 series are labelled in red and blue, respectively.  Other detected lines are mostly Ni
 lines.
 {Alt text: a line plot showing X-ray spectra of Cen X-3 with Resolve.}
}
 \label{fig:broadband}
\end{figure}

\begin{figure}[htpt!]
 \begin{center}
  \includegraphics[width=0.99\columnwidth]{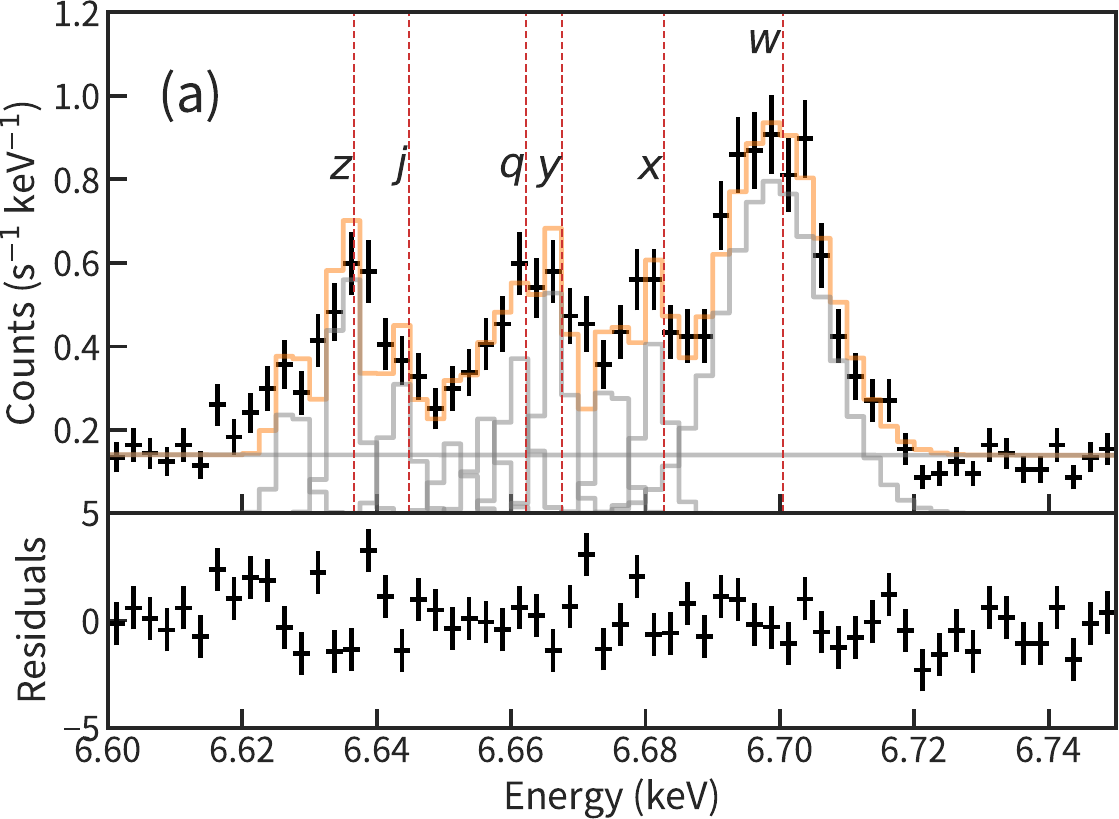}
  \includegraphics[width=0.99\columnwidth]{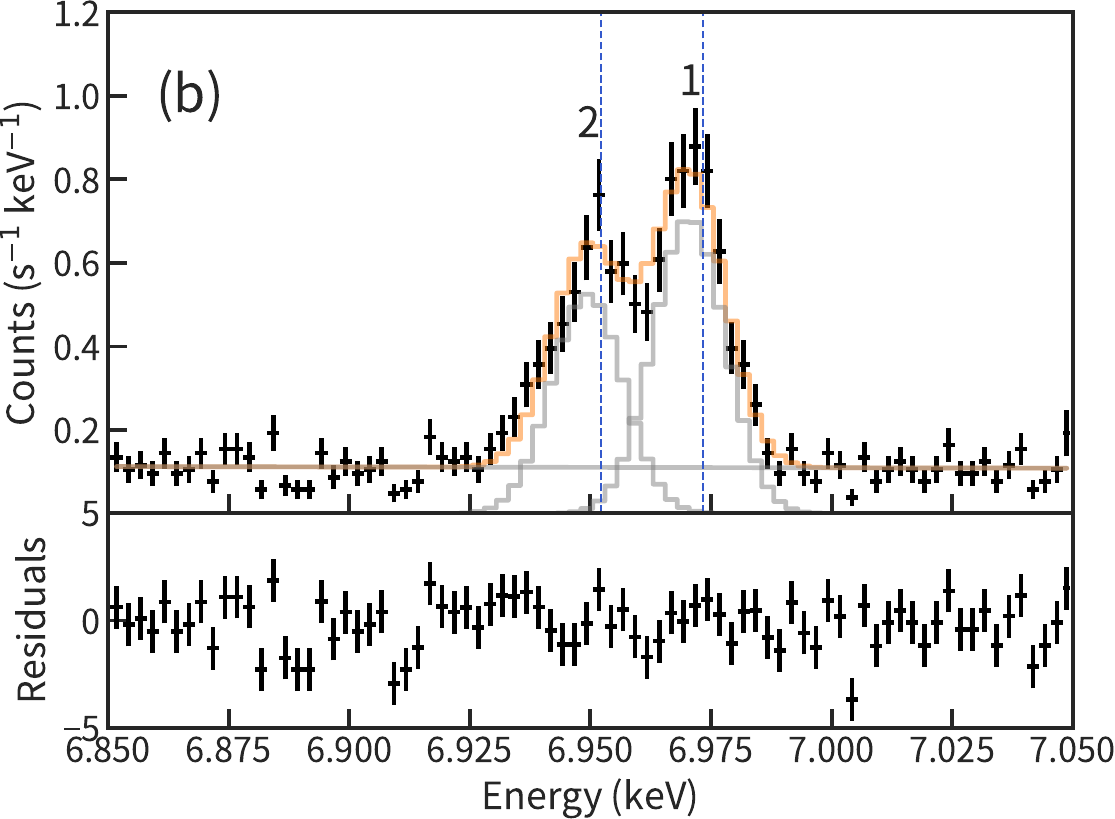}
 \end{center}
 \caption{Results of spectral fitting for the (a) Fe~\emissiontype{XXV} He$\alpha$ and
 (b) Fe~\emissiontype{XXVI} Ly$\alpha$ complexes. The orange lines show the best-fit
 models, and the gray lines are the individual components.
%  The Cash-statistic residuals \citep{kaastra2017} are shown in the bottom panels.
 The bottom panels show the residuals, calculated as $(\text{data} - \text{model})/\text{the data error}$.
 {Alt text: a line plot showing results of spectral fitting.}
}
 \label{fig:fitting}
\end{figure}

Figure \ref{fig:broadband} shows the X-ray spectrum in the Fe K band (6.5--8.75~keV)
during the eclipse. The emission lines of the Fe \emissiontype{XXV} He series
(He$\alpha$, He$\beta$, He$\gamma$, and He$\delta$) as well as those of the Fe
\emissiontype{XXVI} Lyman series (Ly$\alpha$, Ly$\beta$, and Ly$\gamma$) were
detected.
In the He$\alpha$ complex (figure~\ref{fig:fitting}a), the four main lines
($w$, $x$, $y$, $z$) were resolved with some satellite lines. In the Ly$\alpha$ complex
(figure~\ref{fig:fitting}b), the Ly$\alpha_1$ and Ly$\alpha_2$ doublet was
resolved. This is a significant improvement from the HETG spectrum of the same source
(figure~4 in \cite{wojdowski2003}), with which the $\mathcal{R}$ ratio was investigated
earlier.

We characterized these lines by phenomenological fitting.  Within each line complex
(He$\alpha$ and Ly$\alpha$), we modeled the lines with a Gaussian profile on a power-law
continuum to account for the local continuum emission. We considered only the core
Gaussian component in the line spread function for this local fitting.  In the
He$\alpha$ complex, we included the Fe\emissiontype{XXIII}, \emissiontype{XXIV}, and
\emissiontype{XXV} lines whose Einstein A values are $\gtrsim 10^{14}$~s$^{-1}$.  The
included transitions are: $\beta$ (6628.75~eV; 1s2s$^{2}$2p\,$^{1}P_{1}$ $\rightarrow$
2s$^{2}$\,$^{1}S_{0}$), for which the notation follows \cite{doschek1980}; and $a$ (6657.94~eV; 1s2p$^{2}$\,$^{2}P_{3/2}$ $\rightarrow$ 2p\,$^{2}P_{3/2}$),
$k$ (6654.73~eV; 1s2p$^{2}$\,$^{2}D_{3/2}$ $\rightarrow$ 2p\,$^{2}P_{1/2}$), $r$
(6652.94~eV; 1s2s($^{1}S$)2p\,$^{2}P_{1/2}$ $\rightarrow$ 2s\,$^{2}S_{1/2}$), $t$
(6676.23~eV; 1s2s($^{3}S$)2p\,$^{2}P_{1/2}$ $\rightarrow$ 2s\,$^{2}S_{1/2}$), in addition to $w$,
$x$, $y$, $z$, $j$, and $q$ lines, for which the notation follows \cite{gabriel1972}.
In the Ly$\alpha$ complex, we included two lines of the fine-structure doublet
(Ly$\alpha_1$ and Ly$\alpha_2$). The energies of these lines were fixed to the values in
the \texttt{chianti} atomic database version 11.0.2
\citep{feldman1980,briand1984,pike1996,rudolph2013,dere1997,artemyev2005,dere2023}. The
intensities of all lines were fitted individually. The energy shifts were fitted
collectively.  The width of the lines was adjusted collectively for the Ly$\alpha_1$ and
Ly$\alpha_2$ lines in the Ly$\alpha$ complex, while it was done separately for the $w$
line from the others in the He$\alpha$ complex, as only the $w$ line was observed to be
broad.

The results are shown in figure \ref{fig:fitting} and table~\ref{tab:Felines}. The
$\mathcal{R}$ ratio was derived as 0.65 $\pm$ 0.08. The broadening of the $w$ line and
the Ly$\alpha$ complex, which exceeds the natural width, may reflect the velocity
dispersion due to stellar winds as an example, but we put the analysis of the line
profiles out of the scope of this study.

\begin{table}[htpt!]
 \tbl{Fitting results of the lines in the Fe He$\alpha$ and Ly$\alpha$ complex.}{
 \begin{tabular}{lcccc}
  \hline
  Line & Energy & Counts & $v_{\mathrm{broad}}$\footnotemark[$*$] & $v_{\mathrm{shift}}$\footnotemark[$\dagger$]\\
       & (eV)   & (ks$^{-1})$ & (km~s$^{-1}$) & (km~s$^{-1}$)\\
  \hline
   Fe\emissiontype{XXVI} Ly$\alpha_1$ & 6973.24 & 13.2$\pm 0.7$ &  \multirow{2}{*}{788 $^{+44}_{-41}$} & \multirow{2}{*}{$\leq$ 120} \\
   Fe\emissiontype{XXVI} Ly$\alpha_2$ & 6952.13 & 9.8$^{+0.7}_{-0.6}$ & & \\
  \hline
   Fe\emissiontype{XXV} $w$& 6700.40 & 15.5$\pm 0.8$ & 826 $^{+48}_{-45}$ & \multirow{6}{*}{60$\pm 7$} \\
   Fe\emissiontype{XXV} $x$& 6682.70 & 2.1$\pm 0.3$ & \multirow{5}{*}{$\leq$ 1} & \\
   Fe\emissiontype{XXV} $y$& 6667.61 & 2.8$\pm 0.3$ & & \\
   Fe\emissiontype{XXV} $z$& 6636.56 & 3.2$\pm 0.3$ & & \\
   Fe\emissiontype{XXIV} $j$ & 6644.74 & 1.6$\pm 0.3$ & & \\
   Fe\emissiontype{XXIV} $q$ & 6662.24 & 2.0$\pm 0.3$ & &  \\
  \hline
 \end{tabular}
}
  \label{tab:Felines}
 \begin{tabnote}
  \footnotemark[$*$] The line broadening in the full width at half maximum (FWHM).\\
  \footnotemark[$\dagger$] The Doppler shift velocity in line of sight. Positive values
  are for the red-ward shifts.
  \end{tabnote}
\end{table}

\section{Radiative transfer calculation}\label{s4}

We used two RT calculation codes: \texttt{xstar} version 2.59d \citep{kallman2004} and
\texttt{cloudy} release candidate c25
\citep{chakraborty2020a,chakraborty2020,chakraborty2021,chakraborty2022a,chatzikos2023,gunasekera2023}. Both
are two-stream approximation solvers in a one-dimensional setup. For a given incident
radiation field, the balance equation is solved between radiative heating and cooling to
derive the electron temperature ($T_{\mathrm{e}}$) and the rate equations are solved for
the charge balance and level populations, from which emergent spectra are synthesized.

We start by constraining the plasma parameters in subsection~\ref{s4-1} by comparing the
synthesized and observed line ratios. The derived parameters are consistent within
$\lesssim$0.3~dex among the two codes, and we only show \texttt{cloudy} results. In
subsection~\ref{s4-2}, we calculate the $\mathcal{R}_0$ value for the plasma with the derived parameters
and constrain the local UV flux by comparing to the observed $\mathcal{R}$. We present results of
the two codes, as they show some differences, which we also investigate.

\subsection{Estimation of plasma parameters}\label{s4-1}

In our RT calculations, the plasma was represented by a plane parallel slab with the
electron density ($n_{\mathrm{e}}$) and the solar abundances \citep{grevesse1998}. It
has no bulk and minimum (50~km~s$^{-1}$) turbulent velocities. The thickness of the
parallel plane is characterized by the column density ($N_{\mathrm{H}}$). The incident
radiation has a power-law shape with a photon index of 1.3 based on the observed
spectrum (Figure~\ref{fig:broadband}). We confirmed that the result of the $\mathcal{R}$
ratio diagnostic does not depend on the choice of the index in the reasonable range
(1--2) and fixed the value. At the illuminated surface of the slab, the photoionizing
intensity is parameterized by the ionization degree ($\xi_{\mathrm{X}}$). The three
parameters ($n_{\mathrm{e}}$, $N_{\mathrm{H}}$, and $\xi_{\mathrm{X}}$) were evaluated
as follows.

\subsubsection{Electron density ($n_{\mathrm{e}}$)}\label{s4-1-1}

Assuming that the density profile around the O star is governed by its mass loss, we
constrained $n_{\mathrm{e}} < 10^{11}$~cm$^{-3}$ at the position of the NS
\citep{wojdowski2001}. Here, we inferred the density profile assuming the velocity
follows a $\beta$ law \citep{castor1975} with $\beta=0.57$, mass loss rate of $\dot{M} =
1.56 \times 10^{-6}~M_{\odot}~\mathrm{yr}^{-1}$, terminal velocity $v_\infty =
1000$~km~s$^{-1}$, and initial-to-terminal velocity ratio of $v_{0}/v_{\infty} = 0.015$.
The density is far below $n_{\mathrm{crit}} \sim 10^{16}$~cm$^{-3}$ for
Fe\emissiontype{XXV} \citep{blumenthal1972}. The charge and level populations are
predominantly determined by the radiative processes in the photoionized plasmas
(figure~\ref{f01}). As long as $n_{\mathrm{e}} \ll n_{\mathrm{crit}}$ is satisfied, the
interpretation does not depend on the value we choose for the present application, thus
we fixed it to $n_{\mathrm{e}} = 10^{10}$~cm$^{-3}$ for the RT calculation.

\subsubsection{Column density ($N_{\mathrm{H}}$)}\label{s4-1-2}

We use a novel technique to estimate $N_{\mathrm{H}}$ using the Fe\emissiontype{XXVI}
Ly$\alpha$ doublet pair \citep{gunasekera2025}. The intensity ratio of Ly$\alpha_1$
(2p~$^{2}P_{3/2}$ $\rightarrow$ 1s~$^{2}S_{1/2}$) and Ly$\alpha_2$ (2p~$^{2}P_{1/2}$
$\rightarrow$ 1s~$^{2}S_{1/2}$) is 2:1 when the lines are optically thin, following the
statistical weight of their upper levels. When the optical depth increases, the ratio
asymptotes to 1:1 for the ratio of the source function at which the optical depth seen
from the outside is 2/3 in the Eddington-Barbier approximation \citep{barbier1943}. We
should note that Ly$\alpha_3$ (2s~$^{2}S_{1/2}$ $\rightarrow$ 1s~$^{2}S_{1/2}$), which
is indistinguishable from Ly$\alpha_2$, does not contribute much for the line ratio in
this application.

\begin{figure}[htpt!]
 \begin{center}
 \includegraphics[width=1.0\columnwidth]{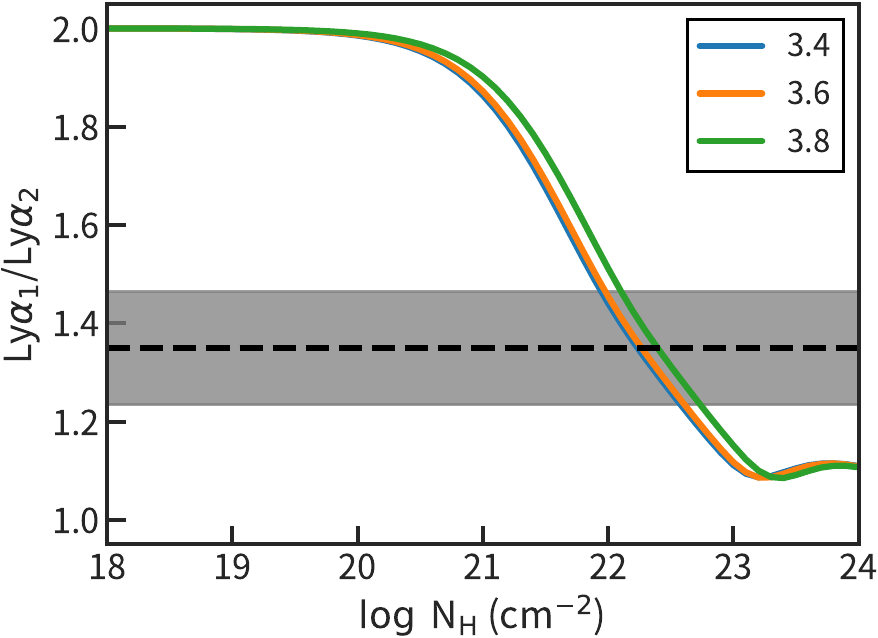}
 \end{center}
 \caption{Fe\emissiontype{XXVI} Ly$\alpha_1$ to Ly$\alpha_2$ line ratio as a function of
 the plasma column density of a plane-parallel slab geometry calculated using
 \texttt{cloudy} for several different $\log \xi$ (erg~cm~s$^{-1}$) values in different
 colors. The observed ratio and its 1 $\sigma$ error range are shown with the black
 dashed line and the gray band.
 Alt text: a line plot showing the observed and calculated Lyman-alpha 1 to 2
 ratio.}
 \label{fig:column-a1a2}
 \end{figure}

Figure~\ref{fig:column-a1a2} shows the RT calculation and the observed ratio
(table~\ref{tab:Felines}) of Ly$\alpha_1$/Ly$\alpha_2$. Their upper levels are
predominantly populated by radiative cascades like $\overline{w}$
(figure~\ref{f01}). The conversion from the line ratio to $N_{\mathrm{H}}$
depends on the charge fraction of Fe$^{26+}$, thus on $\xi$, but only weakly
in our range of interest of $\log \xi = 3.4-3.8$ (erg~cm~s$^{-1}$). Based on this,
we estimate $\log N_{\mathrm{H}} \sim 22$ (cm$^{-2}$).

\subsubsection{Ionization degree ($\xi_{\mathrm{X}}$)}\label{s4-1-3}

Finally, we constrain $\xi$ using the line ratio of the Fe\emissiontype{XXV} He$\alpha$
$w$ and the Fe\emissiontype{XXVI} Ly$\alpha$ line. This reflects the charge fraction
ratio of Fe$^{25+}$ and Fe$^{26+}$, therefore, it sensitively depends on $\xi$
(figure~\ref{f01}). All these lines are optically thick for the estimated
$N_{\mathrm{H}}$ (sub-subsection~\ref{s4-1-2}), which is taken into account in the escape
probability approximation.

\begin{figure}[htpt!]
 \begin{center}
 \includegraphics[width=1.0\columnwidth]{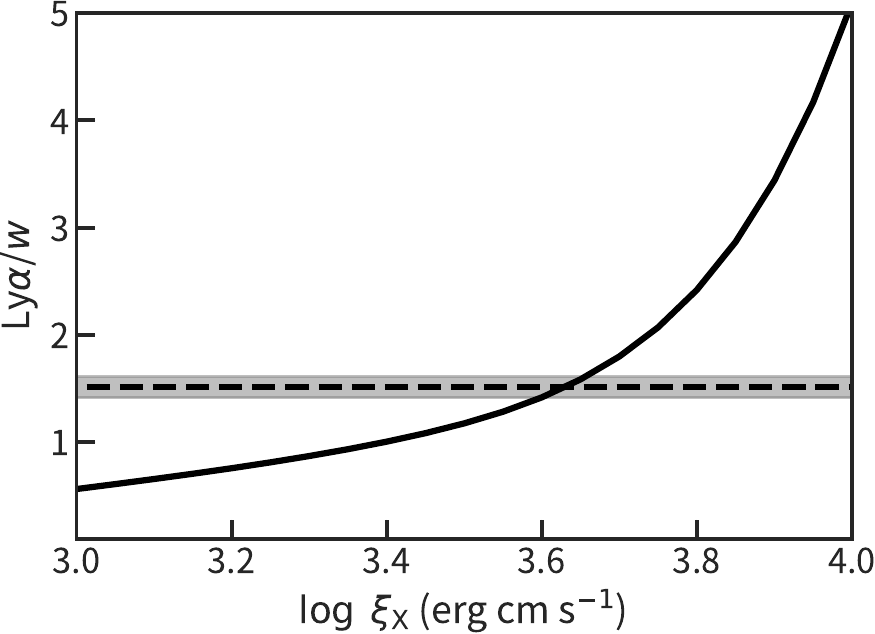}
 \end{center}
 \caption{Fe\emissiontype{XXVI} Ly$\alpha$ to Fe\emissiontype{XXV} $w$ line ratio
 as a function of the ionization degree using \texttt{cloudy} for
 $N_{\mathrm{H}}=10^{22}$~cm$^{-2}$. The doublet (Ly$\alpha_1$ and Ly$\alpha_2$) is
 combined for Ly$\alpha$. The observed ratio and its 1 $\sigma$ error range are shown
 with the black dashed line and the gray band.
 Alt text: a line plot showing the observed and calculated Lyman-alpha to w ratio.}
 \label{fig:ionization}
\end{figure}

Figure~\ref{fig:ionization} shows the RT calculation and the observed ratio
(table~\ref{tab:Felines}) of Ly$\alpha$/$w$. The doublet intensity is summed for
Ly$\alpha$. As $\xi_{\mathrm{X}}$ increases in the range of interest
($\log{\xi_{\mathrm{X}}}=$3--4 erg~cm~s$^{-1}$), the line ratio increases. This is
because the charge fraction of Fe$^{26+}$ increases while that of Fe$^{25+}$ decreases
(figure~\ref{f01}), thus the Ly$\alpha$ intensity increases while the $w$ intensity
decreases. Based on this, we estimate $\log{\xi_{\mathrm{X}}} \sim$3.6
(erg~cm~s$^{-1}$). From the thermal balance equation, the plasma electron temperature is
$T_{\mathrm{e}} = 2.5$~MK.

\subsection{Constraint of the local UV flux}\label{s4-2}

We established the plasma parameters for the photoionized plasma needed to derive
$\mathcal{R}_{0}$. We now compare it with the observed value (subsection~\ref{s3-2}) and
constrain the local UV flux. As discussed in sub-subsection~\ref{s4-1-1}, the $\mathcal{R}$ ratio is
insensitive to density in the regime inferred for this object.

\begin{figure}[htpt!]
 \begin{center}
  \includegraphics[width=1.0\columnwidth]{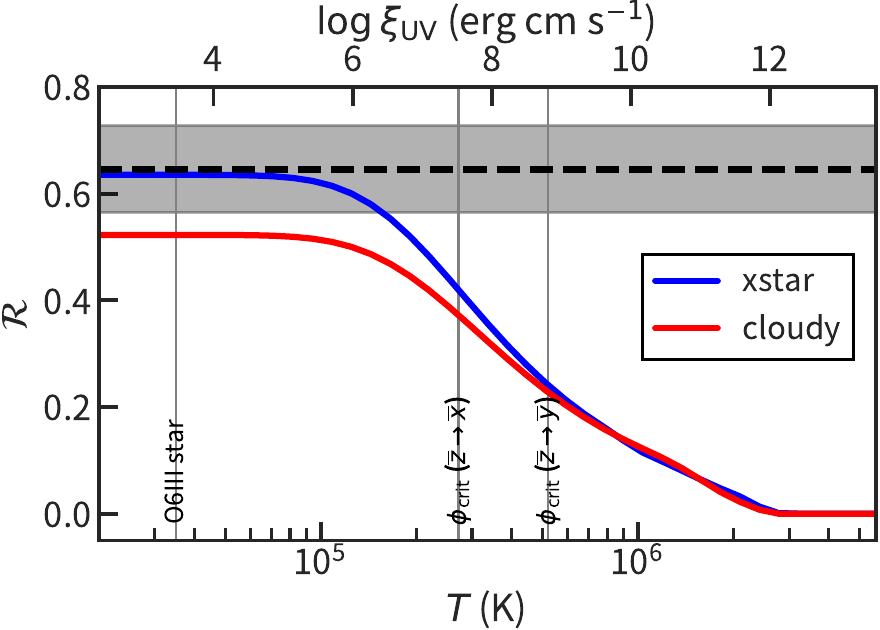}
 \end{center}
 \caption{$\mathcal{R}$ ratio as a function of the local UV flux (top axis) and the
 blackbody temperature (bottom axis; see Figure~\ref{f02}) using \texttt{xstar} and \texttt{cloudy}
 for $N_{\mathrm{H}}=10^{22}$~cm$^{-2}$. The observed
 ratio and its 1 $\sigma$ error range are shown with the black dashed line and the gray
 band. The local UV flux are shown with gray lines corresponding to the blackbody
 temperature of 35~kK (O6III star), 272~kK ($\phi_{\mathrm{crit}}$ for
 the $\overline{z} \rightarrow \overline{x}$ excitation), and 520~kK ($\phi_{\mathrm{crit}}$ for
 the $\overline{z} \rightarrow \overline{y}$ excitation).
 Alt text: a line plot showing the observed and calculated R ratio.}
 \label{fig:Rratio-UVratio}
\end{figure}

Figure \ref{fig:Rratio-UVratio} shows the RT calculation and the observed $\mathcal{R}$
ratio as a function of the local UV flux. Here, the local UV flux is parameterized
similarly with $\xi_{\mathrm{X}}$ as $\xi_{\mathrm{UV}} \equiv
\frac{L_{\mathrm{UV}}}{n_{\mathrm{e}}r^{2}}$, in which $L_{\mathrm{UV}}$ is the
1--1000~Ryd luminosity of the blackbody emission with a temperature of 35~kK for the O
star in Cen X-3 \citep{hutchings1979}. To avoid altering the evaluation by the
additional blackbody component with varying amplitudes, we used $\xi_{\mathrm{X}}$
instead of $\xi_{\mathrm{X}}+\xi_{\mathrm{UV}}$ to parameterize the PIE plasma and fixed
$T_{\mathrm{e}}=2.5$~MK, which is the value obtained without including an additional
blackbody component. As expected, the $\mathcal{R}$ ratio decreases as
$\xi_{\mathrm{UV}}$ increases, with more photons contributing to the photoexcitation
from $\overline{z}$ to $\overline{x}$ and $\overline{y}$.  The calculated $\mathcal{R}$
values are lower for \texttt{cloudy} than for \texttt{xstar}. If we use the
\texttt{xstar} results, the observed $\mathcal{R}$ is consistent with $\mathcal{R}_0$ at
the low $\xi_{\mathrm{UV}}$ limit.

The discrepancy of \texttt{cloudy} from \texttt{xstar} is likely attributable to the
following. To populate $\overline{z}$, the inner-shell photo-ionization of Li-like
Fe$^{+23}$ in the ground level (2s\,$^{2}S_{1/2}$) is important (figure~\ref{f01} right
panel g). In \texttt{cloudy}, this process is included in general, but not in this
particular channel due to the absence of forbidden transitions in the original atomic
database \citep{kaastra1993}.

\begin{figure}[htpt!]
 \begin{center}
  \includegraphics[width=1.0\columnwidth]{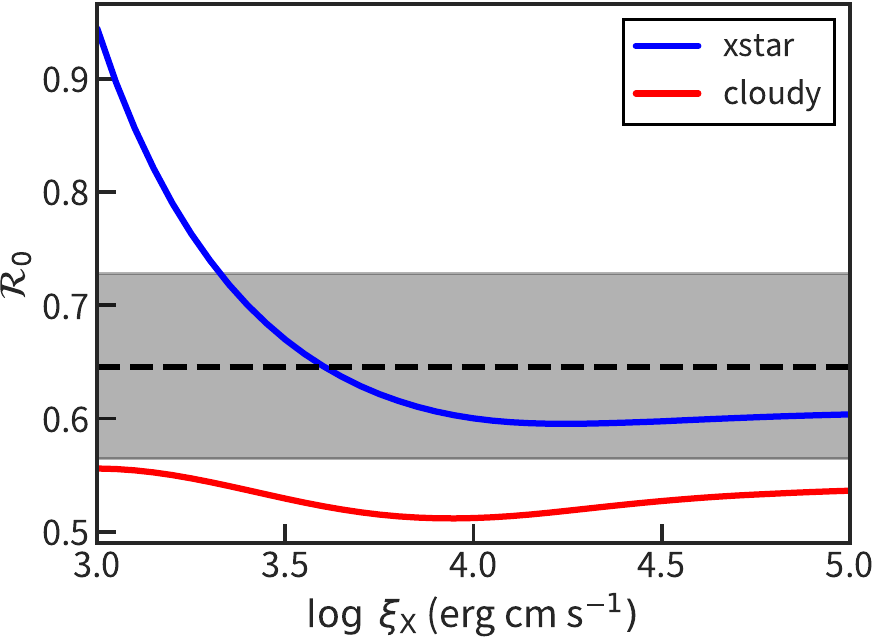}
 \end{center}
 \caption{$\mathcal{R}_{0}$ value as a function of the ionization degree using
 \texttt{xstar} and \texttt{cloudy} for $N_{\mathrm{H}}=10^{22}$~cm$^{-2}$. The observed
 ratio and its 1 $\sigma$ error range are shown with the black dashed line and the gray
 band.
 Alt text: a line plot showing the observed and calculated R-zero value.}
 \label{fig:Rratio-xi}
 \end{figure}

The discrepancy should be clearer at lower $\xi$, at which the inner-shell
photo-ionization is more dominant. Indeed, it is the case in figure
\ref{fig:Rratio-xi}. $\mathcal{R}_{0}$ increases rapidly as $\log{\xi}$ decreases below
$\sim$3.5 due to the increasing contribution of the inner-shell photo-ionization in
\texttt{xstar}, while the same trend is not seen in \texttt{cloudy}. The issue is
isolated in the codes, which will be improved in a future release of \texttt{cloudy}.

\section{Discussion}\label{s5}

\begin{figure}[htpt!]
 \begin{center}
  \includegraphics[width=0.9\columnwidth]{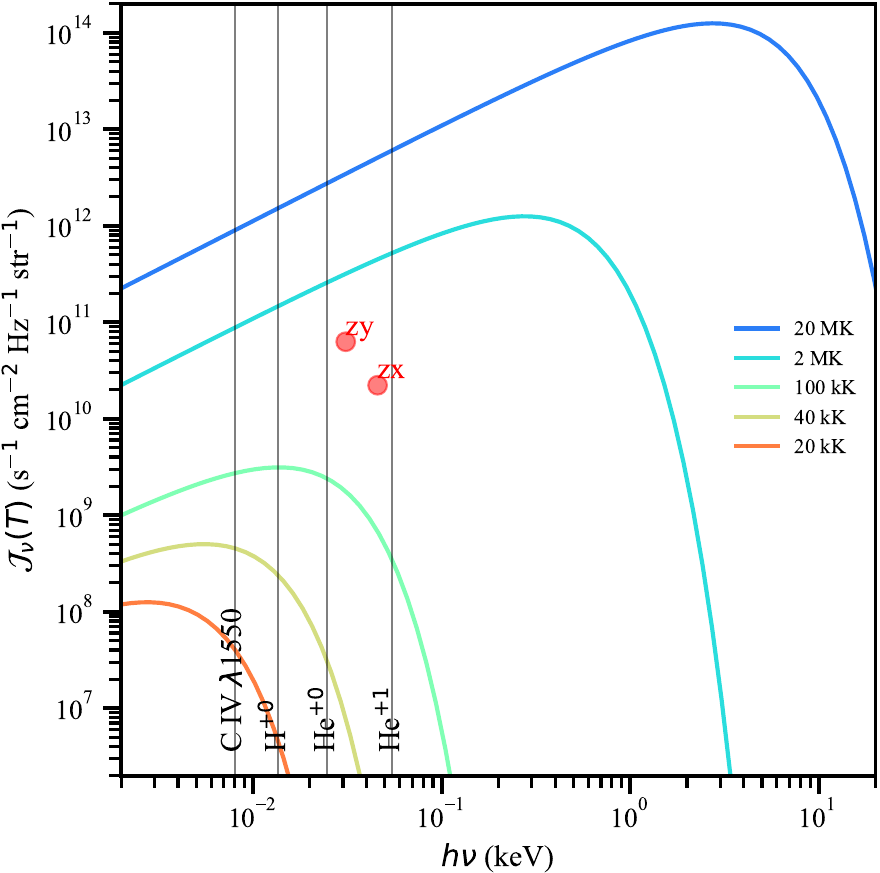}
 \end{center}
 \caption{Specific photon intensity $\mathcal{J}_{\nu}$ in equation~(\ref{e03}) for the
 blackbody at various temperatures ($T$) with solid curves in different colors compared
 to ($h\nu$, $\phi_{\mathrm{crit}}$) of the $\overline{z} \rightarrow \overline{x}$ or
 $\overline{z} \rightarrow \overline{y}$ transitions in red circles. Representative $T$
 ranges are: $T=$20--40~kK for early-type stars (B2 to O3), $T<100$~kK for white dwarfs,
 $T=$2--20~MK for the standard disk \citep{shakura1973} around AGNs with a mass of
 10$^{6}$--10$^{10}$~$M_{\odot}$ and an Eddington ratio of 0.1, and $T=$20~MK for the NS
 surface, or the innermost temperature of the accretion disk of X-ray binaries. The
 energies of the C\emissiontype{IV} line and the ionization potential of H$^{+0}$,
 He$^{+0}$, and He$^{+1}$ are shown with vertical lines.
 Alt text: a line plot showing the blackbody and the critical photon flux values (z to x and y).}
 \label{f02}
 \end{figure}

We first check the validity of the RT calculation. In the microphysics,
$\phi_{\mathrm{crit}}$ is given approximately by
$A_{\mathrm{z}}/B_{\overline{z}\overline{x}}$ or
$A_{\mathrm{z}}/B_{\overline{z}\overline{y}}$, in which $A_{\mathrm{z}}=1.9 \times
10^{8}$~s$^{-1}$ is the Einstein A coefficient from $\overline{z}$ to the ground level,
while $B_{\overline{z}\overline{x}} = 1.2 \times 10^{8}$ and
$B_{\overline{z}\overline{y}} = 6.2 \times 10^{7}$~cm$^{2}$~Hz~str~erg$^{-1}$ are the B
coefficient from $\overline{z}$ to $\overline{x}$ or $\overline{y}$, respectively.  In
the photon flux at the relevant energy ($h\nu=$46 or 31~eV), $\phi_{\mathrm{crit}}/h\nu =
2.3 \times 10^{10}$ or $6.3 \times 10^{10}$~cm$^{-2}$~s$^{-1}$~Hz$^{-1}$~str$^{-1}$ for
$\overline{z}$ to $\overline{x}$ or $\overline{y}$, respectively.  This should be
compared to the specific photon intensity $\mathcal{J_{\nu}}(r)$ at a distance $r$ from
a spherical blackbody with a temperature $T$ and a radius $R$ given by
\begin{equation}
 \label{e03}
  \mathcal{J_{\nu}} = \frac{1}{2} \left[1-\sqrt{1-\left(\frac{R}{r}\right)^2}\right] \frac{2\nu^{2}}{c^{2}} \frac{1}{e^{h\nu/k_{\mathrm{B}}T}-1},
\end{equation}
including the geometrical dilution factor. The $\mathcal{J} (r=R)$ value is given for
several representative $T$ values in figure~\ref{f02}.  The $\phi_{\mathrm{crit}}$ value of the two
transitions ($\overline{z} \rightarrow \overline{x}$ or $\overline{z} \rightarrow
\overline{y}$) is equivalent to the blackbody intensity of $T=272$ or $520$~kK, whose
total luminosity is $L_{\mathrm{UV}} = \sigma_{\mathrm{SB}}T^{4}=3.1 \times 10^{17}$ or
$4.2 \times 10^{18}$~erg~s$^{-1}$~cm$^{-2}$, where $\sigma_{\mathrm{SB}}$ is the
Stefan-Boltzmann constant. This converts to $\xi_{\mathrm{UV}}=3.1 \times 10^{7}$ or
$4.2 \times 10^{8}$~erg~cm~s$^{-1}$ for $n_{\mathrm{e}}=10^{10}$~cm$^{-3}$, which agree
with the value showing $\mathcal{R} \sim 0.5 \mathcal{R}_{0}$ in the RT calculation
(figure~\ref{fig:Rratio-UVratio}).

We found in figure~\ref{fig:Rratio-UVratio} that the observed $\mathcal{R}$ is
consistent with $\mathcal{R}_0$, indicating that the local UV flux $\phi_{\mathrm{UV}}
\ll \phi_{\mathrm{crit}}$. This is expected if the UV source for the photoexcitation is the
O star in Cen X-3. Early-type stars have an effective temperature in the 20--40~kK
range, and none of them produces $\phi_{\mathrm{UV}} > \phi_{\mathrm{crit}}$ for Fe
\emissiontype{XXV} even on the surface. Sufficient UV flux is expected around the
compact object in AGNs and X-ray binaries. The fact that the observed $\mathcal{R} \sim
\mathcal{R}_{0}$ indicates that the X-ray emitting plasma is far away from the NS. From
the geometrical dilution in equation~(\ref{e03}), the distance should be $\gg 10
R_{\mathrm{NS}}$.

\begin{figure}[htpt!]
 \begin{center}
  \includegraphics[width=1.0\columnwidth]{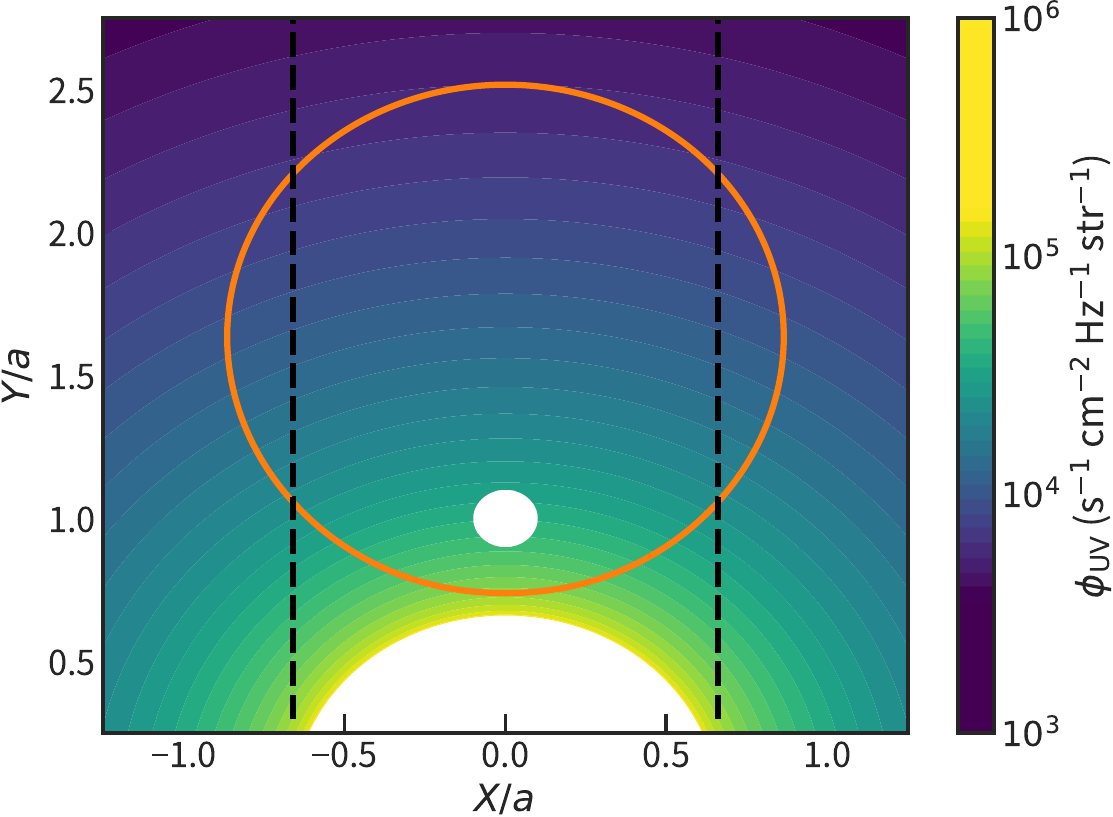}
 \end{center}
 \caption{Heat map of the local UV flux by the O star in Cen X-3 on the orbital plane at
 a mid-eclipse. The coordinates ($X$, $Y$) are given in the unit of the semimajor axis ($a$). The O and NS star is
 placed at (0, 0) and (0, 1), respectively. The size of the NS is not to scale, but the
 others are. The wind parameters are given in sub-subsection~\ref{s4-1-1}. The orange circle indicates
 the region where $\log{\xi_{\mathrm{X}}} = 3.6$ is achieved. The black dashed lines
 represent the boundary of eclipsed part seen from the observer at $(X, Y) = (0,
 -\infty)$.
 Alt text: a color map showing the local UV flux from the O star in Cen X-3.
}
 \label{fig:geometry}
\end{figure}

This is in fact expected from the observed $\xi_{\mathrm{X}}$. By placing the NS in the
stellar wind field of the O star, we can estimate the density of the wind using the wind
parameters in sub-subsection~\ref{s4-1-1}. Figure~\ref{fig:geometry} shows the region, where
$\log{\xi_{\mathrm{X}}} = 3.6$ (erg~cm~s$^{-1}$) is achieved. The region is far from the
NS surface, and the local UV flux is much smaller than $\phi_{\mathrm{crit}}$.

\section{Conclusion}\label{s6}

We obtained high-resolution X-ray spectra of the Fe\emissiontype{XXV} He$\alpha$ and
Fe\emissiontype{XXVI} Ly$\alpha$ complexes from the eclipsed duration of Cen X-3 using
the \textit{Resolve} instrument onboard XRISM. From the line ratio analysis of
Ly$\alpha_1$/Ly$\alpha_2$ and $w$/Ly$\alpha$, we constrained the plasma parameters to be
$\log{\xi_{\mathrm{X}}} \sim 3.6$ (erg~cm~s$^{-1}$) and $\log{N_{\mathrm{H}}} \sim 22$
(cm$^{-2}$). We also derived $\mathcal{R} = 0.65 \pm 0.08$ using the spectra, in which
the four main and several satellite lines were resolved for the first time.

We ran RT calculations for the derived plasma parameters and obtained $\mathcal{R}_{0}$
= 0.65 with \texttt{xstar}.  The smaller value obtained with \texttt{cloudy} is due to
omitted atomic data leading to negation of inner-shell ionization of Fe$^{+23}$ as a
channel to populate the $\overline{z}$ level in the calculation.  The observed
$\mathcal{R}$ is consistent with $\mathcal{R}_{0}$, indicating that there is not
sufficient electron density or UV photon flux to excite from $\overline{z}$ to
$\overline{x}$ and $\overline{y}$ either by electron impact or photo absorption
excitation.

This is indeed consistent with what we expect from the density profile of the O star
wind and the local UV flux from the O star and NS. In this manner, we could verify that
the $\mathcal{R}_{0}$ value calculated by \texttt{xstar} agrees well with the observed
value in the limit of no excitation. This benchmark enables application of $\mathcal{R}$
ratio diagnostics of photoionized plasmas in other systems. In the close vicinity of the
accreting regions of AGNs and X-ray binaries, the UV photon flux is expected to exceed
the $\phi_{\mathrm{crit}}$ value to alter $\mathcal{R}$; this ratio can therefore be used
to probe proximity of X-ray emitting plasmas to accreting matter.

\section*{Acknowledgments}

The authors appreciate Gary J. Ferland at the University of Kentucky, Peter van Hoof at
the Royal Observatory of Belgium, and Stephano Bianchi at Universit\`{a} degli Studi
Roma Tre for clarification of the implementation in \texttt{cloudy}.
This work was supported by the JSPS Core-to-Core Program (grant number:
JPJSCCA20220002). This research made use of the JAXA's high-performance computing system
JSS3. Y\,M is financially supported by the JST SPRING program (grant number: JPMJSP2108)
and JSPS KAKENHI (grant number: JP25KJ0923).
% R\,B
The material is based upon work supported by NASA under award numbers 80GSFC21M0002 and 80GSFC24M0006.
% E\,B
E\,B acknowledges support from NASA grants 80NSSC20K0733, 80NSSC24K1148, and 80NSSC24K1774.
% N\,H
Part of this work was performed under the auspices of the U.S. Department of Energy by Lawrence Livermore National Laboratory under Contract DE-AC52-07NA27344.

\bibliographystyle{aa}
\bibliography{main-20250516}

\end{document}